\documentclass[final,5p,11pt,twocolumn,authoryear]{elsarticle}

\usepackage{graphicx,amsmath,amsfonts,amssymb,bm,mathtools,booktabs,threeparttable,subcaption,nccmath}

\usepackage{hyperref,cleveref}
\hypersetup{colorlinks=true,linkcolor=[rgb]{0.1,0.3,0.9},urlcolor=[rgb]{0.2,0.2,0.2},citecolor=[rgb]{0.1,0.3,0.9}}

\usepackage{color}
\definecolor{myColor1}{rgb}{0.2,0.4,0.8}
\definecolor{myColor2}{rgb}{0.4,0.4,0.4}
\makeatletter
\AtBeginDocument{\def\@linkcolor{myColor1}}
\AtBeginDocument{\def\@urlcolor{myColor2}}
\AtBeginDocument{\def\@citecolor{myColor1}}
\makeatother

\makeatletter\newcommand*\mysizea{\@setfontsize\mysizea{9.5}{11}}\makeatother
\makeatletter\newcommand*\mysizeb{\@setfontsize\mysizeb{8.5}{11}}\makeatother
\makeatletter\newcommand*\mysizec{\@setfontsize\mysizec{7}{11}}\makeatother
\makeatletter\newcommand*\mysized{\@setfontsize\mysized{6.5}{11}}\makeatother

\DeclarePairedDelimiterX{\infdivx}[2]{(}{)}{#1\;\delimsize\|\;#2}
\newcommand{\infdiv}{D_{KL}\infdivx}

\journal{Medical Image Analysis}

\makeatletter
\def\ps@pprintTitle{%
 \let\@oddhead\@empty
 \let\@evenhead\@empty
 \def\@oddfoot{}%
 \let\@evenfoot\@oddfoot}
\makeatother

\begin{document}

\title{FAST-AID Brain: Fast and Accurate Segmentation Tool using Artificial Intelligence Developed for Brain}

\author[add1]{Mostafa Mehdipour Ghazi\corref{cor1}}
\ead{ghazi@di.ku.dk}
\author[add1,add2]{Mads Nielsen}
\ead{madsn@di.ku.dk}
\author{for the Alzheimer's Disease Neuroimaging Initiative\corref{cor2}}

\cortext[cor1]{Corresponding Author.}
\cortext[cor2]{Data used in preparation of this article were obtained from the Alzheimer's Disease Neuroimaging Initiative (ADNI) database (adni.loni.usc.edu). As such, the investigators within the ADNI contributed to the design and implementation of ADNI and/or provided data but did not participate in analysis or writing of this report. A complete listing of ADNI investigators can be found at \url{http://adni.loni.usc.edu/wp-content/uploads/how_to_apply/ADNI_Acknowledgement_List.pdf}}
\address[add1]{Department of Computer Science, University of Copenhagen, Copenhagen, DK}
\address[add2]{Cerebriu A/S, Copenhagen, DK}


\begin{frontmatter}

\begin{abstract}
Medical images used in clinical practice are heterogeneous and not the same quality as scans studied in academic research. Preprocessing breaks down in extreme cases when anatomy, artifacts, or imaging parameters are unusual or protocols are different. Methods robust to these variations are most needed. A novel deep learning method is proposed for fast and accurate segmentation of the human brain into 132 regions. The proposed model uses an efficient U-Net-like network and benefits from the intersection points of different views and hierarchical relations for the fusion of the orthogonal 2D planes and brain labels during the end-to-end training. Weakly supervised learning is deployed to take the advantage of partially labeled data for the whole brain segmentation and estimation of the intracranial volume (ICV). Moreover, data augmentation is used to expand the magnetic resonance imaging (MRI) data by generating realistic brain scans with high variability for robust training of the model while preserving data privacy. The proposed method can be applied to brain MRI data including skull or any other artifacts without preprocessing the images or a drop in performance. Several experiments using different atlases are conducted to evaluate the segmentation performance of the trained model compared to the state-of-the-art, and the results show higher segmentation accuracy and robustness of the proposed model compared to the existing methods across different intra- and inter-domain datasets.
\end{abstract}


\end{frontmatter}

\section{Introduction}

Magnetic resonance imaging (MRI) is a noninvasive imaging modality that can provide high-contrast images with a high spatial resolution. It can represent a detailed morphology of the human brain and be used for structural brain analysis. Therefore, MRI has been widely applied to study the development of neurological diseases and aging through quantitative analysis of the brain MRIs \citep{Yamanakkanavar2020}. The quantitative analysis is usually performed based on volumetric measurements or shape descriptors obtained using accurate segmentation of neuroanatomy of the brain regions \citep{Rashed2020}.

Manual segmentation of structural brain MRI into anatomical regions, i.e., labeling each voxel with a specific tissue type, is an expensive, tedious, and time-consuming process that can be inaccurate due to the shape complexity and human errors or disagreements. Therefore, there is a need for automated segmentation methods that are fast and provide reliable, generalizable, and accurate results. On the other hand, automatic segmentation is very challenging due to the high variability in human anatomy (size, shape, orientation, etc.), image acquisition (settings, contrast, resolution, etc.), and artifacts, and because of the lack of completely annotated data for training.

Atlas-based segmentation tools such as FreeSurfer \citep{Fischl2002}, FSL \citep{Smith2004}, and SPM \citep{Penny2011} use traditional algorithms that typically apply registration to label the structural MRI scans according to manual segmentation. To classify the whole brain, a nonrigid transformation or deformable alignment is used to estimate a transform that spatially transfers the existing dataset to the target domain \citep{Iglesias2015} in a multi-atlas segmentation (MAS) scenario, and the transform is applied to the atlas labels alongside label fusion to form the target labels \citep{Huo2019}. The alignment can be performed in a patch-based manner using a local similarity-based search in the atlas to reduce the computational complexity of the MAS by using a linear alignment \citep{Rousseau2011}. Although MAS methods require less annotated data, they do not generalize well. Moreover, they suffer from high computational costs and deal with ill-conditioned optimization problems.

Data-driven learning-based approaches such as deep neural networks are powerful methods for automatic segmentation of high-dimensional images and extracting functional features for the quantitative analysis of the brain by training a deep model on manually annotated data \citep{Akkus2017}. Although the training and optimization process is time-consuming, the testing or prediction procedure can be done quite fast. Previous studies have successfully applied convolutional neural networks (CNNs) to the brain MRI segmentation problem using a patch-based or semantic examination \citep{Bernal2019}. Compared to the full convolutional training in which the network uses the 2D slices or the whole 3D volumes as the input, patch-based networks are trained on local neighborhoods of the input scans. Therefore, depending on the used patch size, patch-based networks are more prone to learn intensity-based features and less able to represent the shape \citep{Wachinger2018}. In addition, these networks can suffer from the class imbalance and difference between the patches which can cause trouble in the network convergence (fluctuation) as they may require more training iterations to cover all available classes in the image while facing many background samples.

The existing memory issues in processing MRI volumes cause segmentation methods such the SLANT \citep{Huo2019} to mostly be performed in a patch-based manner with sliding windows. This limits the network receptive field and requires the network to be trained from scratch. The 2D segmentation methods apply 2D CNNs to single slices and are fast in learning easy components or differences (e.g., tumors) in the image. However, they lack 3D contextual information from adjacent slices and result in discontinuous predictions, leading to failure in more challenging tasks. To address this problem, 2.5D methods were proposed that modify 2D architectures to incorporate 3D information. For instance, QuickNAT \citep{Roy2019} and FastSurfer \citep{Henschel2020} applied 2D CNNs to the three principal views (axial, coronal, and sagittal) and aggregated the results using fixed voting weights to infer the final segmentation. Similarly, multiple 2D views \citep{Roth2014} or thickened 2D inputs using adjacent slices \citep{Yu2019} were fed to 2D CNNs as different channels of the input. Alternatively, recurrent neural networks (RNNs) were applied to an ordered series of 2D slices for segmentation using 2D CNN-based architectures \citep{Chen2016,Poudel2016}. Finally, combinations of patch-based 3D CNNs and 2D CNNs were successfully applied to MRI segmentation \citep{Mehta2017,Isensee2021}. However, the existing 2.D methods do not actually learn 3D representations during the training, and the improvements are mostly made by fusing the individually learned networks' predictions.

Brain segmentation using deep learning typically faces a major problem with generalization due to the scarcity of annotated data (privacy, difficulty, and cost), high anatomical and structural variability of human brains (age, gender, ethnicity, and health condition), and difference in MRI acquisition (scanner device, strength, resolution, contrast, and artifacts) \citep{Krupa2015}. To cope with the imaging artifacts, many preprocessing and correction steps have been used in the literature before the final segmentation. Also, registration techniques are used to compensate for the high variability of the brains. However, CNNs require a lot of diverse data for training to ensure the robustness of the trained model. Therefore, data augmentation is suggested as an efficient technique for increasing the training samples by applying plausible modifications to the available data. For example, synthetic intensity inhomogeneity can randomly be added to brain images to show the effectiveness of data augmentation against preprocessing \citep{Khalili2019}.

\begin{figure*}[!t]
\centering
\includegraphics[scale=0.98]{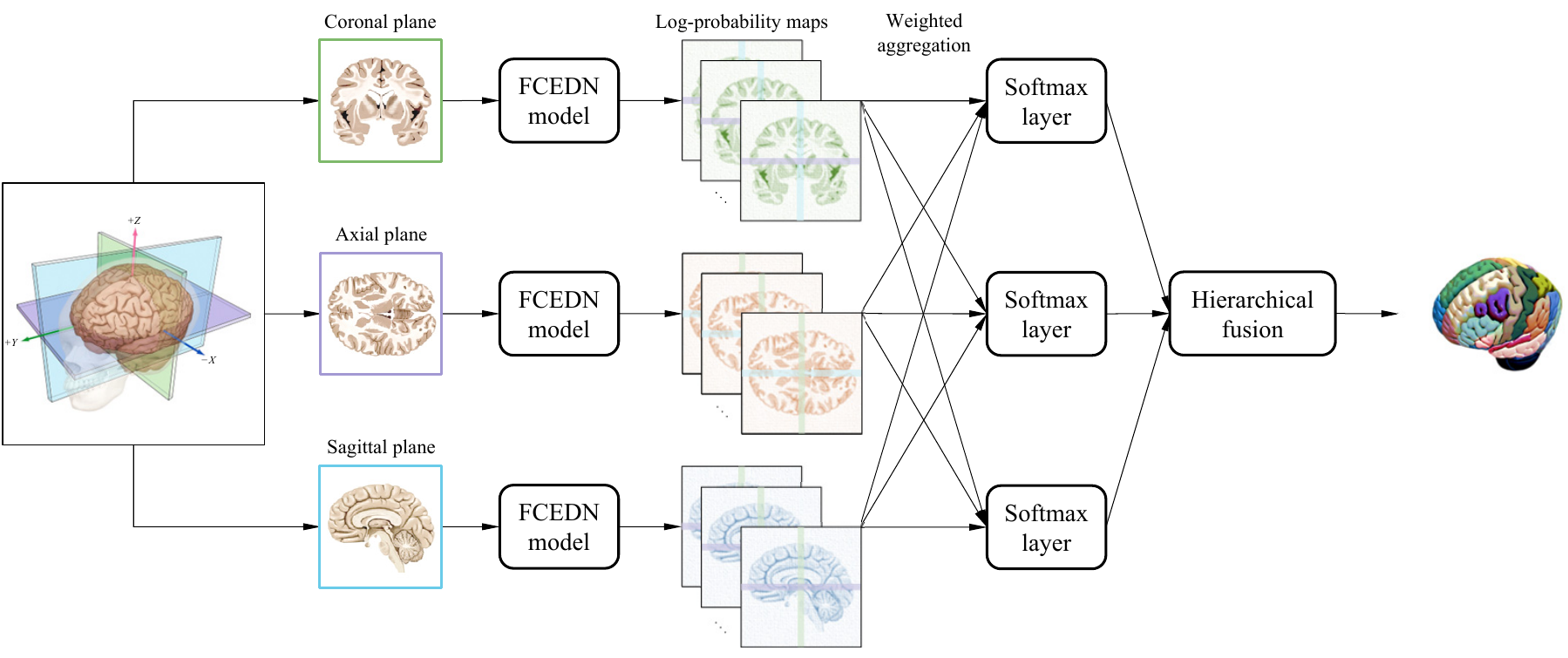}
\caption{The proposed model for brain parcellation. The model takes a RAS-oriented, T1-weighted brain MRI scan as its input and segments it into 132 cortical and noncortical regions. The network applies a trained FCEDN model to the three orthogonal planes, aggregates the output maps using the learned planar weights from the intersection points for 3D volume reconstruction, and uses the hierarchical parcellation for voxel classification.}
\label{fig_net}
\end{figure*}

Another concern in automatic brain segmentation is the accuracy of the predictions and parcellation, especially for the detailed areas of the brain such as gyri and sulci \citep{Sendra2020}. The high variability of the brain structures in size and shape leads to a highly imbalanced classification problem which hinders the application of the classic losses, e.g., based on cross-entropy (CE) \citep{Murphy2012} or Dice similarity coefficient (DSC) \citep{Dice1945}, as the network will be prone to overfit the larger regions and ignore the smaller ones. To tackle the problem, different modifications have been applied to the loss functions mostly by using class prior related weights \citep{Lin2017,Sudre2017,Salehi2017,Abraham2019,Cui2019}. Hierarchical relations between the classes can also be taken into account to effectively address the issue and improve the parcellation performance \citep{Redmon2017,Zhang2017,Hu2018,Muller2020,Graham2020}. This can facilitate network training and testing on multiple datasets with various degrees of label granularity \citep{Demyanov2017}.

To this end, we propose an efficient 2.5D-based deep learning method for automatic segmentation of the human brain into 132 cortical and noncortical regions with an average intra/inter-domain DSC of 0.75 in less than 40 seconds on GPU. The proposed network applies a U-Net-like fully convolutional network to the three principal views and learns to efficiently fuse them based on the intersection points and hierarchical relations in an end-to-end training fashion. The main contributions of this work are fivefold; first, we propose a novel deep learning method that can train on 2D slices while effectively incorporating 3D information and compare it with the state-of-the-art 2.5D-3D deep learning methods applied to the same datasets; second, we use label hierarchies to handle the class imbalance issue and improve the parcellation accuracy; third, we use weak supervision to learn from partially labeled data to segment the whole brain and estimate the intracranial volume (ICV); fourth, we simulate several MRI artifacts and augment the training data on-the-fly to improve the robustness and generalizability and address privacy-preserving learning problem; fifth, we conduct exhaustive experiments on many different atlases to evaluate the accuracy and robustness of the trained model for brain segmentation and the stability of the estimated ICVs compared to the state-of-the-art.

\section{Methods}

\subsection{The Proposed Model}

\subsubsection{Segmentation Network}

The proposed network trains a fully convolutional encoder-decoder network (FCEDN) on three orthogonal planes (axial, coronal, and sagittal) and benefits from the intersection points of the planes for 3D volume reconstruction. The network uses only one backbone to encode representations from different planes which keeps the complexity of the network low while enabling it to incorporate the 3D information. Figure \ref{fig_net} shows the proposed model for brain volume segmentation. As can be seen, the three perpendicular planes are individually fed to the same U-Net-like network to obtain the class-membership scores per plane before applying the Softmax to the weighted sum scores.

\subsubsection{Hierarchical Softmax}

\begin{figure*}[!t]
\centering
\includegraphics[scale=0.98]{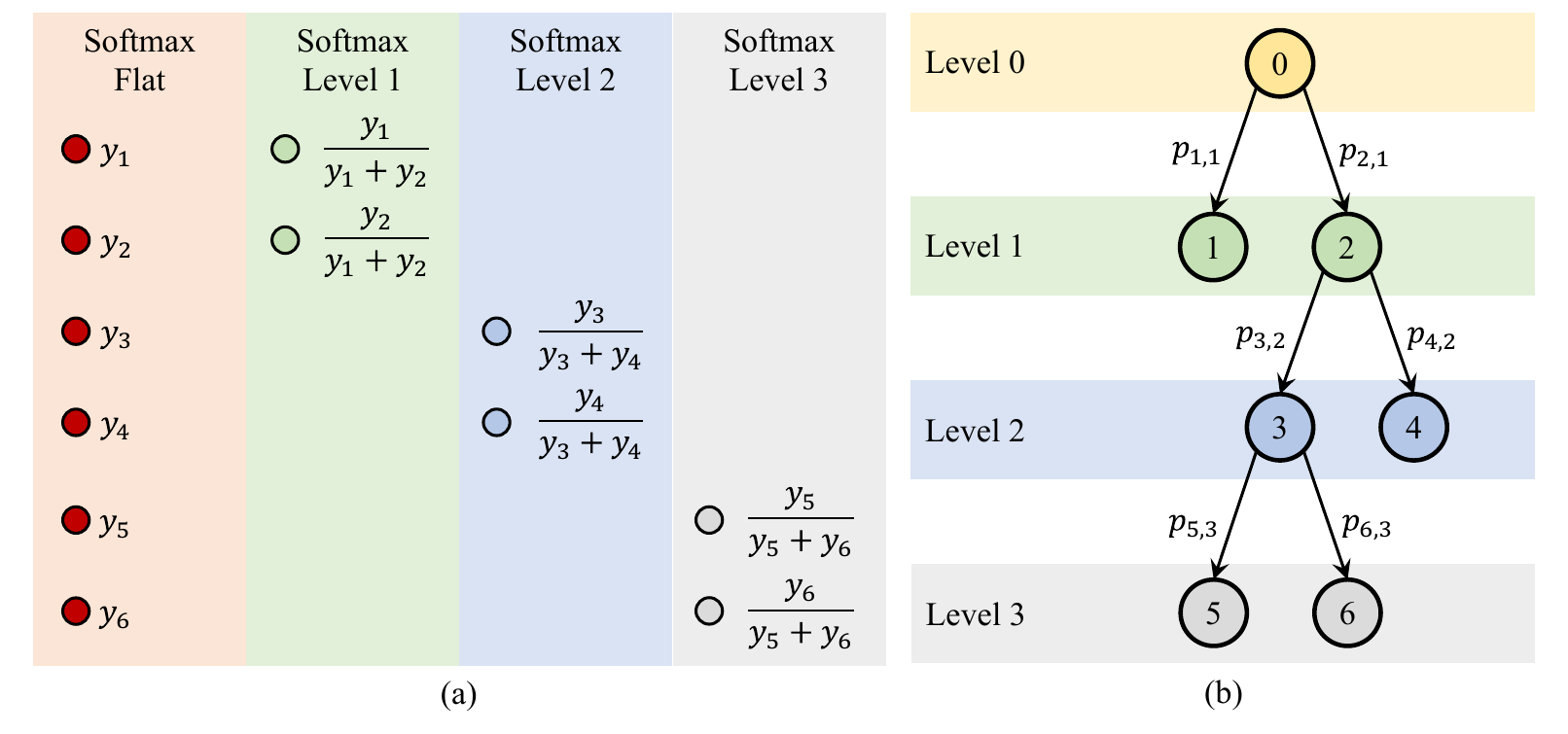}
\caption{An illustration of how the hierarchical Softmax scores are calculated per pixel (a) using a sample tree (b). Note that the tree includes three levels and six label nodes (leaves). For instance, the output probability for the node labeled as $4$ at the second level can be obtained as $p(4) = p(4 | 2) p(2)$, where $p(4 | 2) = p_{4,2} = y_4 / (y_3 + y_4)$ and $p(2) = p(2 | 0) p(0)$, where $p(2 | 0) = p_{2,1} = y_2 / (y_1 + y_2)$ and $p(0) = 1$. Note that the sum of the conditional probabilities on the branches of each node is equal to 1.}
\label{fig_tree}
\end{figure*}

We apply a hierarchical parcellation method to pixel classification using the Softmax classifier mentioned by \citet{Redmon2017}; in contrast to flat parcellation in which output nodes of the Softmax layer indicate the class membership probabilities, the hierarchical Softmax scores denote the conditional probabilities where output probabilities at each level nodes are conditioned to their previous level parent nodes. These scores are obtained by normalizing the flat Softmax scores based on the available sibling nodes per level so that branches of each level node sum to one. If $n_{i,l}$ indicates the child node or branch $i$ at level $l$, the normalized Softmax score of this node is calculated as the conditional probability $p(n_{i,l} | n_{i,l-1}) = y_{i} / \sum_j y_{j}$, where $n_{i, l-1}$ denotes the $i$'s parent node, $y_{i}$ is the flat Softmax score of the node $i$, and $j$ spans the node $i$ and its sibling nodes at level $l$. Therefore, the class-conditional probability $p(n_{i,l})$ assigned for the child node can be determined by the product of the obtained level score and its parent scores at previous levels up to the root as
\begin{fleqn}\label{eq_fuse}
\begin{equation}
\begin{split}
p(n_{i,l}) & = p(n_{i,l}, n_{i,l-1}) = p(n_{i,l} | n_{i,l-1}) p(n_{i,l-1}) \\
& = p(n_{i,l} | n_{i,l-1}) p(n_{i,l-1} | n_{i,l-2}) \dots p(n_r) \,,
\end{split}
\end{equation}
\end{fleqn}
\noindent where $n_r$ is the root node and its probability is one. Figure \ref{fig_tree} shows how the hierarchical Softmax scores are calculated using a sample tree including three levels and six label nodes. Finally, the hierarchical CE loss for a sample pixel $m$ is calculated based on the obtained class-conditional probabilities for each level node as
\begin{fleqn}\label{eq_celoss}
\begin{equation}
\textstyle \mathcal{L}(m) = - \sum_{l} \sum_{j \in l} t_{m}(n_{j,l}) \log p_{m}(n_{j,l}) \,,
\end{equation}
\end{fleqn}
\noindent where $t_{m}(n_{j,l})$ is a one-hot encoded array indicating the true association of the pixel $m$ to the label node $j$.

\subsubsection{3D Fusion}

Before the Softmax function is applied to the output scores of the planes and the CE loss is calculated, the intersection point of the planes is updated based on a weighted aggregation of the three corresponding points (log probabilities) per class. This is a learning-based alternative to the correlated probability fusion method \citep{OBrien1999} which reaps the benefits of 3D segmentation while sidestepping its computational challenges. Additionally, the consistency between intersection lines of the pairs of the planes is determined based on the average Kullback-Leibler divergence (KLD) \citep{Kullback1951} between the corresponding points of every two orthogonal planes and is added to the total loss. Finally, the overall loss is obtained by the accumulation of the hierarchical losses averaged across all available pixels of each plane and the consistency terms as
\begin{fleqn}
\begin{equation*}
\begin{split}
\mathcal{L}_{total} & = \frac{1}{\mathcal{M}} \sum_{m=1}^{\mathcal{M}} (\mathcal{L}_A(m) + \mathcal{L}_C(m) + \mathcal{L}_S(m)) \\
& + D(AC, CA) + D(CS, SC) + D(AS, SA) \,,
\end{split}
\end{equation*}
\end{fleqn}
\noindent where $A$, $C$, and $S$ refer to the axial, coronal, and sagittal planes, respectively, and $\mathcal{M}$ is the number of available pixels per plane. Besides, the three additional terms measure the distance between the pixels of the intersection line of each pair of the planes using the average KLD metric defined as
\begin{fleqn}
\begin{equation*}
\textstyle D(\bm{P}, \bm{Q}) = \frac{1}{2}\big(\infdiv{\bm{P}}{\bm{Q}} + \infdiv{\bm{Q}}{\bm{P}}\big) \,,
\end{equation*}
\end{fleqn}
\noindent where $\bm{P}$ and $\bm{Q}$ are two arbitrary arrays ($\mathcal{K}$ pixels by $\mathcal{C}$ classes each) and $D_{KL}$ is the (asymmetric) KLD operator applied as
\begin{fleqn}
\begin{equation*}
\begin{split}
\infdiv{\bm{P}}{\bm{Q}} = \frac{1}{\mathcal{K}} \sum_{k=1}^{\mathcal{K}} \sum_{c=1}^{\mathcal{C}} \bm{P}(k,c) \log \Big(\frac{\bm{P}(k,c)}{\bm{Q}(k,c)}\Big) \,, \\
\infdiv{\bm{Q}}{\bm{P}} = \frac{1}{\mathcal{K}} \sum_{k=1}^{\mathcal{K}} \sum_{c=1}^{\mathcal{C}} \bm{Q}(k,c) \log \Big(\frac{\bm{Q}(k,c)}{\bm{P}(k,c)}\Big) \,.
\end{split}
\end{equation*}
\end{fleqn}

\subsubsection{Weak Supervision}

The available training data is partially annotated, where some of the labels are missing from different scans, some of which can be ignored or fused for consistent training. However, there are some scans with brain voxels labeled as the cranial cavities, which can be used for the estimation of the intracranial volume (ICV), an important normalization measure used to correct for head size in studies associated with brain volume changes. Since the hierarchical loss cannot cover the missing cranial cavities surrounding the whole brain region, we propose using a weakly supervised learning approach inspired by \citet{Ronneberger2015,Nguyen2020}. More specifically, the background voxels, including the missing cavity labels, are assigned uncertainty weights calculated based on a Gaussian kernel with the Euclidean distance transform ($d$) of the foreground voxels as
\begin{fleqn}
\begin{equation*}
w = e ^ {- d ^ 2 / 2 \sigma ^ 2} \,,
\end{equation*}
\end{fleqn}
\noindent where $\sigma$ is the standard deviation or bandwidth of the kernel and controls the width of the area surrounding the brain and is set to $\sqrt{10}$ mm, estimated based on the scans with available cranial cavity labels. Finally, the one-hot encoded array $t$ mentioned in \eqref{eq_celoss} is replaced with the uncertainty weights for the voxels with missing labels and background classes.

\subsubsection{Prediction Approach}

Test prediction can be performed in two different ways. The first approach is to apply the learned model to the brain slices from different views and stack them to obtain log-probability maps of the three score volumes. These 4D arrays can then be combined using the learned aggregation weights and fed into the Softmax layer for hierarchical label fusion using \eqref{eq_fuse}. The final classification is obtained by classifying each pixel based on the node associated with the highest probability score. This approach requires large memory for storing 4D arrays (3D MRI volume size times the number of hierarchical level nodes). The alternative approach is to use the majority voting algorithm \citep{Littlestone1994} which is rather fast and requires less memory. In this manner, we obtain the three planar label volumes, classified without combining the scores, and make the final decision based on the majority of the labels per pixel. In the case where no majority label is found per pixel, the label associated with the higher aggregation weight is selected.

\begin{figure*}[!t]
\centering
\includegraphics[scale=0.375]{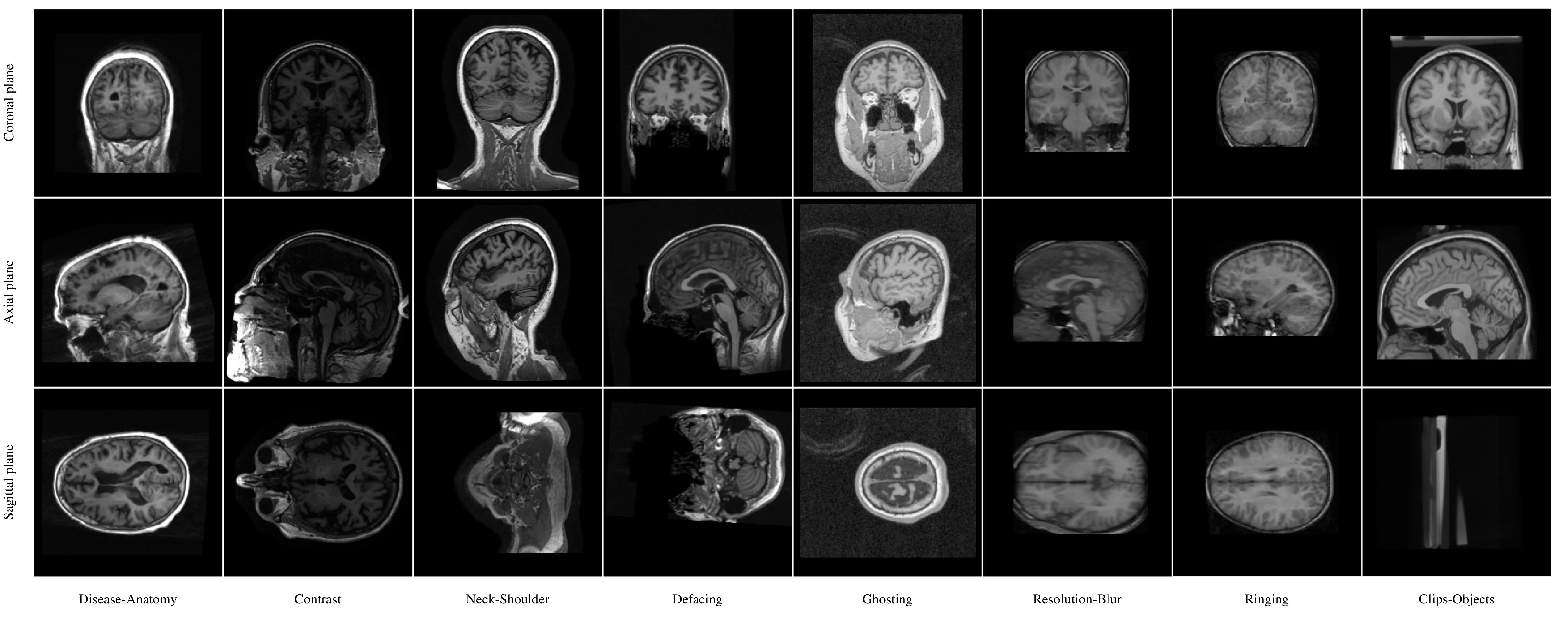}
\caption{Multiview illustration of the heterogeneous brain MRI data used in this study. The high variability of the scan parameters indicates the importance of data augmentation for training robust models on the data expanded with different realistic artifacts and changes.}
\label{fig_bad_samples}
\end{figure*}

\subsection{Data Preprocessing and Augmentation}

In this section, all the utilized preprocessing and augmentation techniques are described. We have mainly focused on data augmentation and attempted to use very few preprocessing steps to accelerate the training convergence and testing phase and make the models robust to domain shift, which can drop the generalization accuracy due to the differences in the test data (see Figure \ref{fig_bad_samples}). All of the utilized transforms are 3D and fast and can be applied on the fly during training to the original scans, making the network eligible for online learning. The source codes of the developed tool for generating realistic distortions or deformations on the MRI scans are made publicly available at \url{https://github.com/Mostafa-Ghazi/MRI-Augmentation}.

\subsubsection{Head Orientation Correction}

Using a standard head orientation for all brain scans can improve the training and prediction performance for brain segmentation. RAS orientation is a neurologically preferred convention where the coordinate system in $+X$, $+Y$, $+Z$ directions, as depicted in Figure \ref{fig_net}, is oriented towards the right, anterior, and superior of the head, respectively. We apply the affine matrix stored in headers of the scans to transform all brain positions in the RAS orientation \citep{Shen2014}.

\subsubsection{Resolution Adjustment}

Since different scans can have different isotropic/anisotropic voxel sizes, in order to train the brain segmentation network, we may need to resize the volume dimensions considering the spacing or slice thicknesses. Therefore, we scale the brain volumes to the nearest even integers, after multiplying the original array dimensions by the corresponding spacing. The resampling can be done by using the linear and nearest-neighbor interpolations for the image and label volumes, respectively. Later, we pad or crop the scanned brain volumes by adding or removing constant/background slices around the brains to obtain scans of the same dimension 256$\times$256$\times$256 with an isotropic spacing of 1 mm.

\subsubsection{Contrast Adjustment}

Contrast adjustment is an image processing technique that remaps pixel intensities to a stretched display range by sharpening differences between low and high pixel values. To this end, we normalize the intensity values of the volumes to [0, 1] using the available dynamic range of each scan, which saturates the high and low intensity values and stretches the distribution to fill the entire intensity range. In addition, we apply the gamma transform \citep{Chen2018Gamma} to the corrected image volumes with random values in [0.8, 1.2] to augment data with slightly different contrast.

\subsubsection{Volume Rotation}

The orientation of the brain can be slightly different for various scans even after the head position correction. Therefore, we augment the available volumes with randomly rotated ones by allowing the volumes to be rotated about the three perpendicular coordinate axes with an angle randomly chosen in [$-$10$^\circ$, 10$^\circ$]. Linear and nearest-neighbor interpolations are applied to the image and label volumes, respectively.

\subsubsection{Skull Stripping and Defacing}

Skull stripping \citep{Iglesias2011} and defacing \citep{Theyers2021} are preprocessing steps that aim at removing facial features or more areas surrounding the brain. These techniques can help with brain extraction and address clinical data privacy issues by securely training models on anonymized data. Hence, to enable the network to focus on learning representations from the brain regardless of the presence of the face, skull, ears, neck, or shoulders, we extract volumes of randomly cropped areas encompassing the brain.

\subsubsection{Noise Addition and Multiplication}

The MRI images are usually prone to suffer from additive and multiplicative noises such as Gaussian and speckle \citep{Ali2018}. Hence, to make the output predictions robust to these types of noises and improve the generalization accuracy, we augment the available intensity volumes with distorted ones by introducing a zero-mean Gaussian noise or a speckle noise with variances randomly chosen in [0, 0.0001] mm$^2$.

\subsubsection{Intensity Inhomogeneity Distortion}

Intensity variation or nonuniformity across the image is a common problem in MRI acquisition and can be due to several reasons such as the failure of the radio frequency coil, induced eddy currents, B1 field inhomogeneity, and scanning nonferromagnetic materials \citep{Erasmus2004}. Although preprocessing techniques have been used to estimate the bias field to remove intensity inhomogeneity \citep{Sled1998}, deep learning methods can obtain superior segmentation results with data augmentation using synthetically introduced intensity inhomogeneity \citep{Khalili2019}. On this account, we train the network using data containing simulated intensity inhomogeneities by multiplying an elliptic gradient field with the brain volumes \citep{Hui2010}. Assuming images of the same cubic dimension 256, the gradient field is calculated based on the equation of an ellipse in standard form using the points from a structured rectangular grid with integer values from 1 to 256, centers randomly chosen in [1, 256], and radiuses of 256.

\subsubsection{Ringing Artifact Augmentation}

The ringing disturbance is a Gibbs phenomenon that occurs as oscillation at boundaries with high contrast transitions. It is caused by under-sampling or truncation of high-frequency components in the image \citep{Erasmus2004}. We augment this artifact by applying the centralized fast Fourier transform (FFT) to the brain volumes in three orthogonal directions and cutting the edges of the k-space \citep{Moratal2008} at a random integer in [90, 120] along the three axes.

\subsubsection{Ghosting Artifact Augmentation}

The ghosting noise is a phase-encoded motion that appears as repeated versions of the scanned object in the image \citep{Erasmus2004}. It is caused by periodic movements of tissue or fluid during the scan, affecting data sampling in the phase-encoding direction. We augment this artifact by modulating the k-space lines of each axis differently; we weight every $n$-th component of the k-space (FFT) per dimension by a random factor in [0.85, 0.95], where $n$ is a random integer in [2, 4], representing the number of the repeated brains.

\subsubsection{Elastic Deformation}

Elastic distortion is a state-of-the-art method \citep{Simard2003} for expanding the training data by synthesizing plausible transformations of data, and hence, learning shape-invariant representations. Accordingly, we apply the random elastic deformation algorithm to augment our training data. First, a random 3D uniform displacement field is generated along each axis. The obtained random fields are smoothed using a Gaussian filter with an elasticity coefficient $\sigma$ randomly chosen in [20, 30] and a square kernel size of $2\lceil2\sigma\rceil+1$. They are scaled then with a factor $\alpha$ randomly selected in [200, 500], which controls the intensity of the deformation. Finally, a structured rectangular grid with integer values from 1 to 256 is interpolated with the MRI volume to obtain a plausibly deformed volume. Linear and nearest-neighbor interpolations are applied to the image and label volumes, respectively.

\section{Experimental Setup}

\subsection{Data}

The core data used in this study contains 107 T1-weighted MRI volumes in NIFTI format with manual annotations obtained from Neuromorphometrics, Inc. (\url{http://www.neuromorphometrics.com/}). The labels are assigned for each voxel by highly trained neuroanatomical technicians using the NVM tool \citep{Worth2001} and indicate the neuroanatomical structure present at the voxel. The exact specification of the labels is defined based on Neuromorphometrics’ general segmentation protocol (\url{http://neuromorphometrics.com/Seg/}) and its cortical parcellation protocol \citep{Tourville2010}. 

In addition to the abovementioned similarly annotated datasets, five separate datasets including 183 differently annotated and around 500 unannotated brain MRI scans are used to assess the across-cohort generalizability of the trained models to the unseen data. The annotated datasets include fewer segmented brain areas and are nearly matched with our standard atlas labels in the overlapping regions.

The T1-weighted MRI images are obtained using different scanners (Siemens, GE, and Philips) from the Open Access Series of Imaging Studies (\url{ http://www.oasis-brains.org/}), the Centre for the Developing Brain (\url{http://brain-development.org/}), the Center for Morphometric Analysis at Massachusetts General Hospital (\url{https://mail.nmr.mgh.harvard.edu/mailman/listinfo/ibsr}), and the Alzheimer’s Disease Neuroimaging Initiative (\url{http://adni.loni.usc.edu/data-samples/access-data/}). The ADNI was launched in 2003 as a public-private partnership, led by principal investigator Michael W. Weiner, MD. The primary goal of ADNI has been to test whether serial MRI, PET, other biological markers, and clinical and neuropsychological assessment can be combined to measure the progression of mild cognitive impairment and early Alzheimer's disease.

\begin{figure*}[!t]
\centering
\includegraphics[width=0.92\linewidth,height=0.98\textheight]{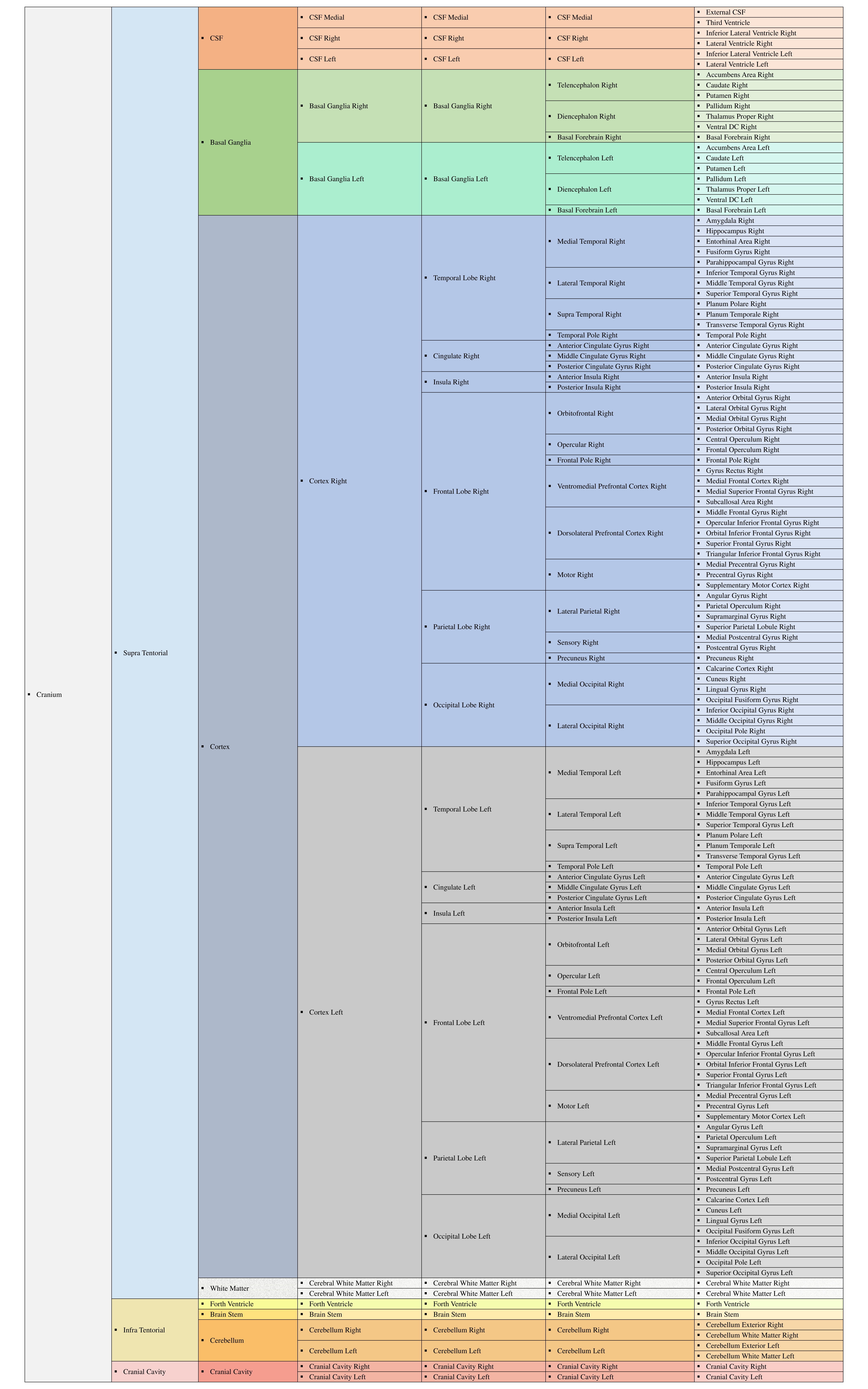}
\caption{The hierarchical tree used for label fusion and loss calculations.}
\label{fig_labels}
\end{figure*}

We use 132 brain labels (95 cortical and 37 noncortical) as well as the background for brain segmentation. These labels involve four unpaired structures for basic segmentation and 64 symmetric structures for cortical parcellation and are obtained after excluding rare annotations and matching labels from different cohorts. Finally, the labels are hierarchically fused in seven levels using a hierarchical label tree with four unpaired structures (third ventricle, fourth ventricle, brain stem, and CSF) and 64 anatomical structures as shown in Figure \ref{fig_labels}.

\subsubsection{20Repeats}

This dataset is collected from the first phase of the OASIS \citep{Marcus2007} and contains 40 MRI scans from 20 normal subjects (12 females and 8 males) aged between 19 and 34. The brain MRI scans are obtained from a 1.5T scanner at an isotropic resolution of 1 mm per voxel with the annotations provided by the Neuromorphometrics.

\subsubsection{MICCAI}

This dataset is collected from the first phase of the OASIS \citep{Marcus2007}, annotated by the Neuromorphometrics, and used in the Medical Image Computing and Computer-Assisted Intervention 2012 Multi-Atlas Labeling Challenge \citep{Landman2012} for subcortical structure segmentation. It contains 35 MRI scans from 35 normal subjects (22 females and 13 males) aged between 18 and 90. The brain MRI scans are obtained from a 1.5T scanner at an isotropic resolution of 1 mm per voxel.

\subsubsection{Demo}

This sample MRI scan is obtained from a 19-year-old male subject in the first phase of the OASIS \citep{Marcus2007} using a 1.5T scanner at an isotropic resolution of 1 mm per voxel with the annotations provided by the Neuromorphometrics, which are available online at \url{http://www.neuromorphometrics.com/1103_3.tgz}.

\subsubsection{Colin27}

This dataset is collected at the McConnell Brain Imaging Centre \citep{Holmes1998} (\url{http://www.bic.mni.mcgill.ca/ServicesAtlases/Colin27Highres}) and contains an average volume of 27 scans from the same normal subject obtained at an isotropic spatial resolution of 0.5 mm with the annotations provided by the Neuromorphometrics.

\subsubsection{ADNI30}

This dataset is collected from the first phase of the ADNI study \citep{Jack2008} and contains 30 MRI scans (15 demented and 15 elderly controls) acquired from 29 patients (15 males and 14 females), aged between 62 and 88. It contains both 1.5T and 3T scans obtained at an average resolution of 1 mm $\times$ 1 mm $\times$ 1.2 mm per voxel with the annotations provided by the Neuromorphometrics.

\subsubsection{HarP}

This dataset is collected from the ADNI study \citep{Jack2008} and contains 135 MRI scans of normal, cognitively impaired, and demented subjects (70 males and 65 females) aged between 60 and 90. It contains both 1.5T and 3T scans obtained at an isotropic resolution of 1 mm per voxel with the hippocampal segmentations provided based on the EADC-ADNI Harmonized Hippocampal Protocol \citep{Boccardi2015} (\url{http://www.hippocampal-protocol.net/SOPs/index.php}).

\subsubsection{Hammers}

This data consists of 95 manually delineated regions drawn on MRI scans of 30 healthy subjects (15 males and 15 females) aged between 20 and 54, having no neurological, medical, or psychiatric conditions. The dataset was acquired using a 1.5T scanner placed at the epilepsy MRI unit \citep{Faillenot2017} (\url{https://soundray.org/hammers-n30r95/}) at an average isotropic spacing of 0.94 mm per voxel.

\subsubsection{IBSR}

The scans were obtained from 18 healthy subjects (14 males and 4 females) aged between 7 and 71, at an average resolution of 0.94 mm $\times$ 0.94 mm $\times$ 1.5 mm, with the manual segmentations of 34 regions provided by the Internet Brain Segmentation Repository \citep{Rohlfing2011} (\url{https://www.nitrc.org/projects/ibsr/}).

\subsubsection{ADNI1-2Yr}

This dataset is obtained from the standardized datasets of the ADNI \citep{Wyman2013} named as ADNI1:Complete 2Yr 1.5T (\url{https://adni.loni.usc.edu/methods/mri-tool/standardized-mri-data-sets/}), and it contains 1.5T MRI scans of 503 subjects at their baseline and possibly 12-month and 24-month follow-ups obtained at an average resolution of 1 mm $\times$ 1 mm $\times$ 1.2 mm per voxel. The subjects are normal, cognitively impaired, or demented (292 males and 211 females) and aged between 55 and 90.

\subsection{Hyperparameters}

Before training our final model for brain segmentation, we optimize the initial parameter values of the network using the Bayesian Optimization algorithm \citep{Snoek2012}. We select a set of optimization parameters including the learning rate and weight decay in $[10^{-6}, 10^{-4}]$ and the core network using DeepLab V3+ CNN \citep{Chen2018} with the base networks of ResNet-18 \citep{He2016}, ResNet-50 \citep{He2016}, and MobileNetV2 \citep{Sandler2018}, all trained on ImageNet (\url{http://www.image-net.org}), as well as U-Net \citep{Ronneberger2015}, and SegNet \citep{Badrinarayanan2017}. Applying a 5-fold cross-validation on a subset of the annotated data suggests that the DeepLab V3+ CNN with ResNet-50 as the backbone network using a base learning rate and a weight decay of $5\times10^{-5}$ can result in the highest segmentation accuracy amongst others.

The training subset of the MICCAI challenge dataset (15 scans), the second scan of the subjects from the 20Repeats dataset (20 scans), and 25 randomly selected scans from the ADNI30 were used for training while the demo sample was applied to validation. The adaptive moment estimation (Adam) algorithm \citep{Kingma2014} was used to update the network parameter values with a minibatch size of 4 orthogonal slices, a gradient decay factor of 0.9, a squared gradient decay factor of 0.99, and a base learning rate and an L2-norm regularization factor of $5\times10^{-5}$. The networks were trained for at most 136,000 iterations using the early-stopping method with 10,880 iterations of patience and a piecewise learning rate schedule with a drop factor of 0.9 per 5,440 iterations.

\subsection{Evaluation Metrics}

The $F_{1}$-measure, also known as the Dice similarity coefficient (DSC) \citep{Dice1945}, is used to gauge the similarity of the predicted and true segmentation results based on the number of overlapping pixels in both sets as
\begin{fleqn}
\begin{equation*}
\mathrm{DSC}(Y, T) = \frac{2 |Y \cap T|}{|Y| + |T|} \,,
\end{equation*}
\end{fleqn}
\noindent where $|Y|$ and $|T|$ represent the cardinal numbers of the predicted and true label sets of the region of interest. The intersection indicates the true positives, and the denominator terms sum over the false positives, false negatives, and twice the true positives.

Besides the spatial similarities evaluated through the DSC, we use a volumetric similarity metric \citep{Taha2015} which considers the absolute difference between the volume size of the true and segmented regions as
\begin{fleqn}
\begin{equation*}
\mathrm{VS}(Y, T) = 1 - \frac{||Y| - |T||}{|Y| + |T|} \,,
\end{equation*}
\end{fleqn}
\noindent where the absolute volume difference can be seen as the difference between the false positives and false negatives.

Finally, to see how the obtained results from different models on various datasets are statistically significantly different, we use the two-sided Wilcoxon signed-rank test \citep{Wilcoxon1945}.

\subsection{Testing Strategy}

We use different state-of-the-art deep learning architectures and tools including nnU-Net \citep{Isensee2021}, Multi-Planar U-Net \citep{Perslev2019}, and FastSurfer \citep{Henschel2020} to compare the segmentation results. To be more specific, nnU-Net is the state-of-the-art medical image segmentation architecture developed based on U-Net \citep{Ronneberger2015} and FastSurfer has shown to be the most accurate and robust model available for brain MRI segmentation. In comparison to the SLANT \citep{Huo2019} which heavily uses preprocessing and postprocessing techniques together with a patch-based 3D CNN-based segmentation method, FastSurfer takes the advantage of the traditional segmentation tool of FreeSurfer \citep{Fischl2002} while being very fast and making the software publicly available for clinical-research use. Accordingly, we have also made our tool publicly available at \url{https://github.com/Mostafa-Ghazi/FAST-AID-Brain}.

We use different datasets to compare the intra- and inter-domain segmentation accuracy of the models. Statistical tests are used to compare the significance level of the difference between the obtained results. In addition, we discuss the segmentation goodness by showing the accuracy on the 20Repeats dataset where 20 subjects were scanned twice and independently annotated by clinical experts \citep{Worth2015}. Finally, we estimate the ICV using the proposed method for different datasets and show its stability in comparison with the state-of-the-art \citep{Malone2015,Sargolzaei2015}.

\section{Results and Discussion}

\begin{table*}[!t]
\centering
\small
\caption{The test DSC (mean$\pm$SD) of the different models trained on the annotated MICCAI challenge dataset for segmenting the brains into 130 regions. The best results are highlighted in boldface.}
\label{table1}
\renewcommand{\arraystretch}{1.6}
\centering
\begin{tabular}{cccc}
\toprule
nnU-Net 2D & nnU-Net 3D & Multi-Planar U-Net & FAST-AID Brain \\
\citep{Isensee2021} & \citep{Isensee2021} & \citep{Perslev2019} &  \\
\bottomrule
0.723$\pm$0.185 & 0.763$\pm$0.138 & 0.735$\pm$0.144 & \textbf{0.771}$\pm$\textbf{0.122} \\
\toprule
\end{tabular}
\end{table*}

\begin{table*}[!t]
\centering
\mysizea
\caption{The test segmentation accuracy (mean$\pm$SD) of the different trained models on the different annotated test sets. The best results are highlighted in boldface and are statistically significantly different ($p < 0.05$). Note that the models with different labels are evaluated based on the common regions, indicated as “*”.}
\label{table2}
\renewcommand{\arraystretch}{1.6}
\centering
\begin{tabular}{lcccccc}
\toprule
Dataset & \multicolumn{2}{c}{20Repeats} & \multicolumn{2}{c}{MICCAI} & \multicolumn{2}{c}{ADNI30} \\
Metric & DSC & VS & DSC & VS & DSC & VS \\
\bottomrule
FAST-AID Brain & 0.788$\pm$0.085 & 0.941$\pm$0.062 & 0.752$\pm$0.128 & 0.916$\pm$0.094 & 0.707$\pm$0.135 & 0.901$\pm$0.101 \\
FAST-AID Brain (*) & \textbf{0.797}$\pm$\textbf{0.126} & \textbf{0.948}$\pm$\textbf{0.056} & \textbf{0.835}$\pm$\textbf{0.077} & \textbf{0.952}$\pm$\textbf{0.047} & \textbf{0.798}$\pm$\textbf{0.085} & \textbf{0.947}$\pm$\textbf{0.046} \\
FastSurfer \citep{Henschel2020} (*) & 0.639$\pm$0.140 & 0.885$\pm$0.099 & 0.710$\pm$0.141 & 0.875$\pm$0.102 & 0.615$\pm$0.170 & 0.875$\pm$0.102 \\
\toprule
\end{tabular}
\end{table*}

\begin{table*}[!t]
\centering
\mysizec
\caption{The segmentation accuracy (mean$\pm$SD) of the different trained models on the inter-domain test sets annotated differently. The best results are highlighted in boldface and are statistically significantly different ($p < 0.05$). Note that the models with different labels are evaluated based on the common regions, indicated as “*”.}
\label{table3}
\renewcommand{\arraystretch}{1.6}
\centering
\begin{tabular}{lcccccccc}
\toprule
Dataset & \multicolumn{2}{c}{Colin27} & \multicolumn{2}{c}{HarP} & \multicolumn{2}{c}{Hammers} & \multicolumn{2}{c}{IBSR} \\
Metric & DSC & VS & DSC & VS & DSC & VS & DSC & VS \\
\bottomrule
FAST-AID Brain (*) & \textbf{0.792}$\pm$\textbf{0.128} & \textbf{0.901}$\pm$\textbf{0.102} & 0.718$\pm$\textbf{0.064} & \textbf{0.965}$\pm$\textbf{0.031} & \textbf{0.543}$\pm$\textbf{0.037} & \textbf{0.926}$\pm$\textbf{0.049} & \textbf{0.765}$\pm$\textbf{0.032} & \textbf{0.948}$\pm$\textbf{0.044} \\
FastSurfer \citep{Henschel2020} (*) & 0.632$\pm$0.145 & 0.870$\pm$0.103 & \textbf{0.729}$\pm$0.072 & 0.880$\pm$0.043 & 0.522$\pm$0.192 & 0.721$\pm$0.177 & 0.757$\pm$0.150  & 0.901$\pm$0.141 \\
\toprule
\end{tabular}
\end{table*}

\subsection{Intra-Domain Generalization}

In the first experiment, we examine the segmentation accuracy of the trained models on the annotated test set from the same domain. Table \ref{table1} compares the DSC of the different state-of-the-art methods on the annotated test set of the MICCAI challenge. As it can be seen, our proposed model achieves the best segmentation accuracy compared to the alternatives.

Next, we assess the segmentation accuracy of the proposed model trained on the core annotated datasets compared to that of the FastSurfer \citep{Henschel2020}. The segmentation results from the different annotated test sets are presented in Table \ref{table2} for both models. To have a fair comparison between the two models, we calculate the accuracy based on the common regions excluding the different labels. As can be seen, the proposed model achieves higher accuracy compared to FastSurfer in all datasets. One explanation for the accuracy difference could be that FastSurfer is trained on noisy annotations, obtained from FreeSurfer \citep{Fischl2002}, while using three separate 2D models in a flat classification scenario. It should also be noted that FastSrufer is trained on rather big data (140 vs. 60 scans) including the OASIS and ADNI cohorts and spans various anatomical and acquisition parameters. Moreover, in both cases, the segmentation precision sees a drop in the ADNI30 cohort. This could be because the other datasets contain preprocessed scans with better quality or higher resolution and more samples available for training.

\begin{figure*}[!t]
\centering
\begin{subfigure}[t]{0.98\textwidth}
\raisebox{-\height}{\includegraphics[scale=0.3]{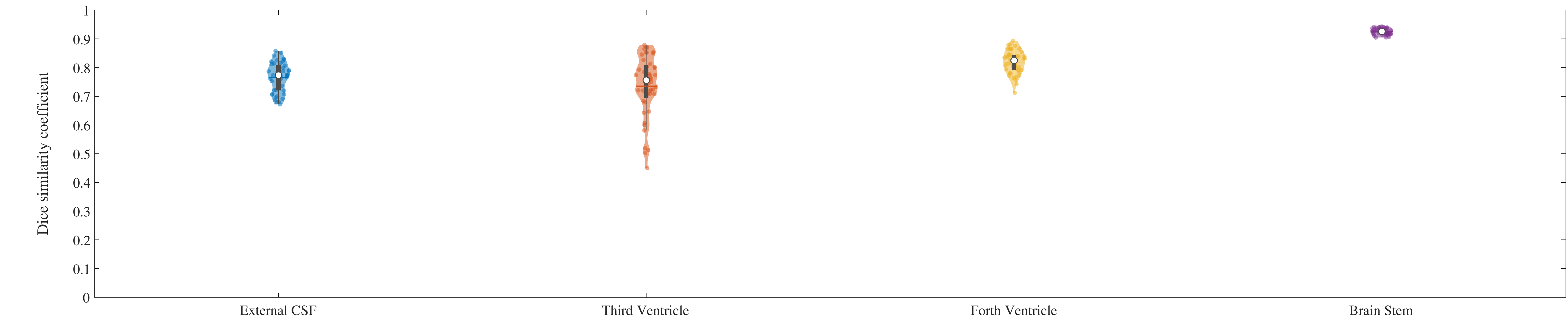}}
\end{subfigure}
\begin{subfigure}[t]{0.98\textwidth}
\raisebox{-\height}{\includegraphics[scale=0.3]{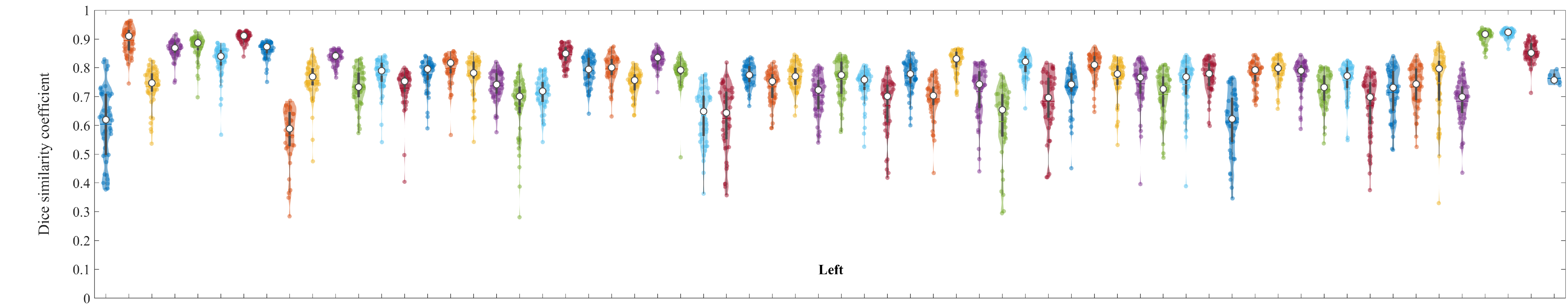}}
\end{subfigure}
\begin{subfigure}[t]{0.98\textwidth}
\raisebox{-\height}{\includegraphics[scale=0.3]{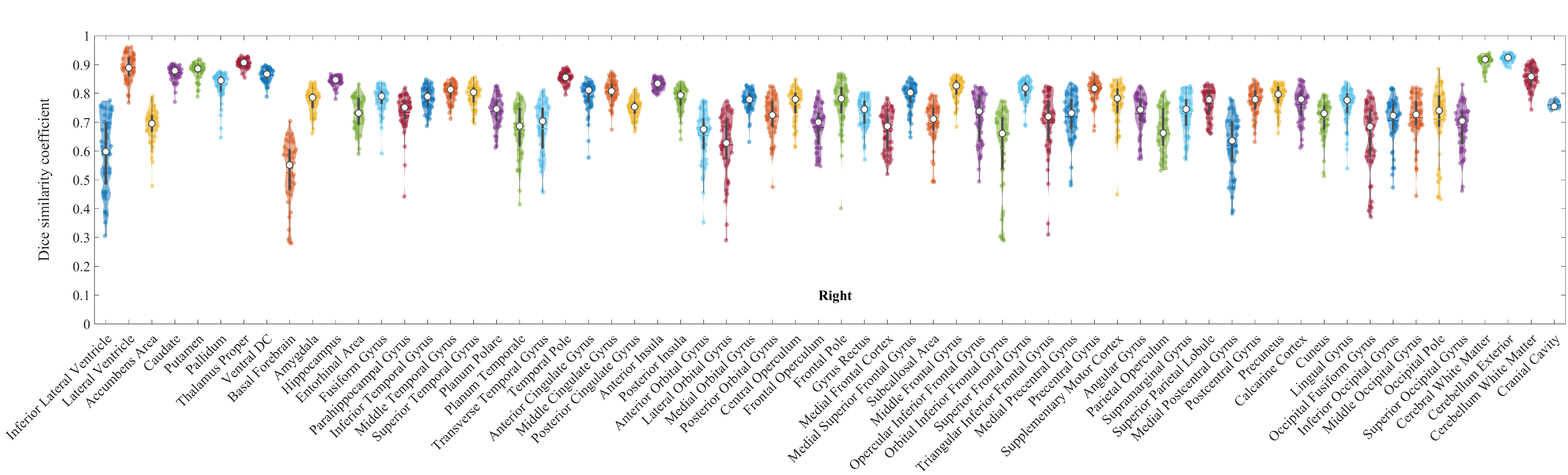}}
\end{subfigure}
\caption{The violin plots of the Dice scores for 132 segmented brain regions using FAST-AID Brain on different intra-domain test sets, i.e., 20Repeats, MICCAI, and ADNI30. The three plots from the top represent the asymmetric, left, and right regions of the brain, respectively. The encompassed white circles show the median points on the violin plots, and the transparent areas visualize the kernel density plots or distributions of the scattered points.}
\label{fig_fastaid_intra}
\end{figure*}

\begin{figure*}[!t]
\centering
\begin{subfigure}[t]{0.98\textwidth}
\raisebox{-\height}{\includegraphics[scale=0.3]{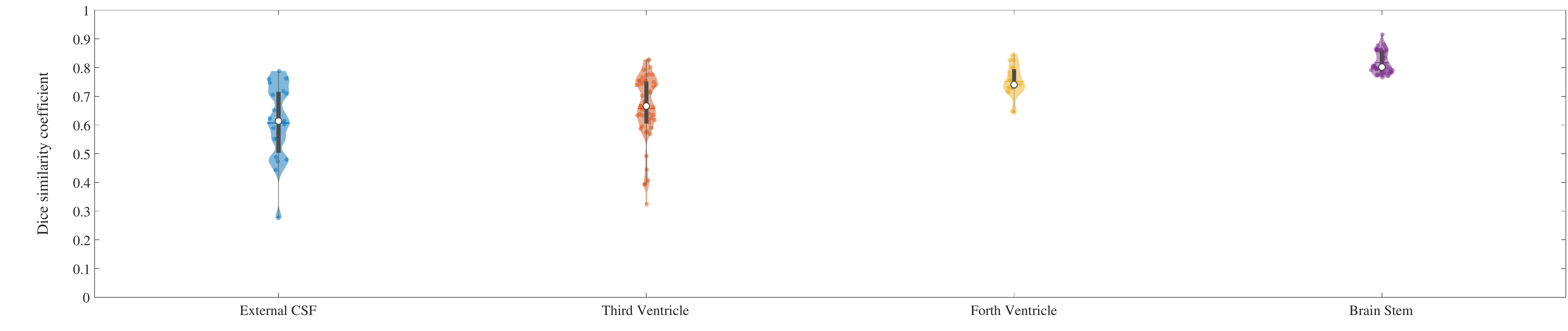}}
\end{subfigure}
\begin{subfigure}[t]{0.98\textwidth}
\raisebox{-\height}{\includegraphics[scale=0.3]{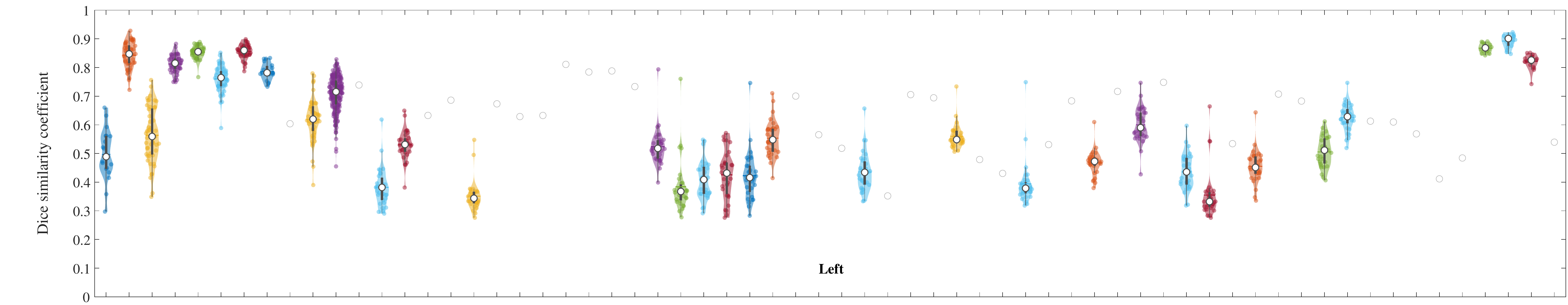}}
\end{subfigure}
\begin{subfigure}[t]{0.98\textwidth}
\raisebox{-\height}{\includegraphics[scale=0.3]{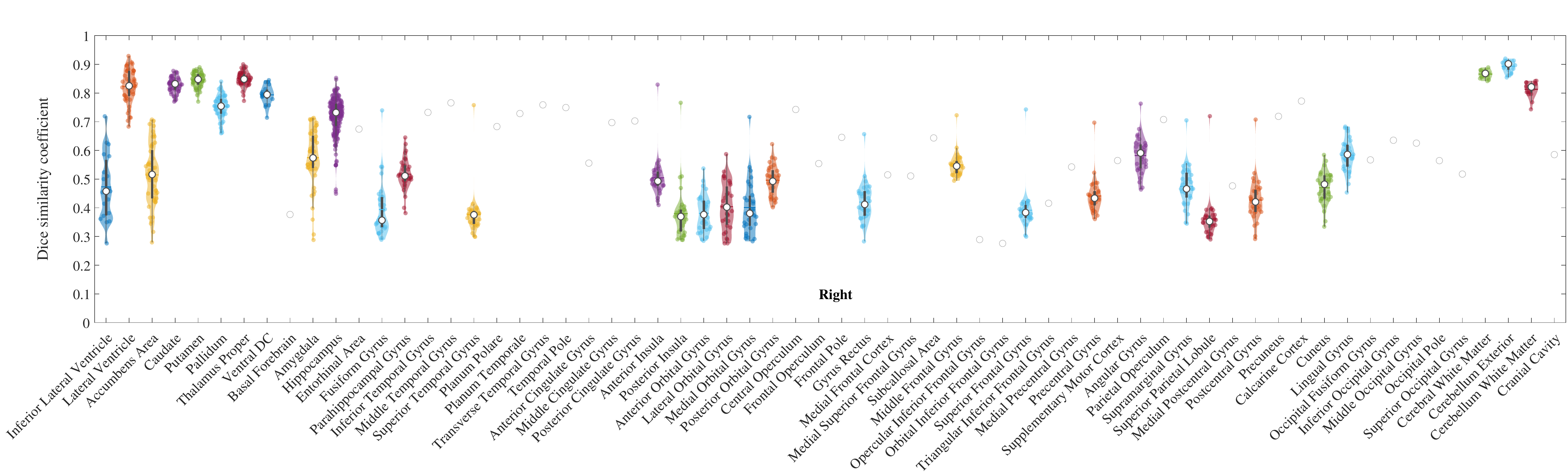}}
\end{subfigure}
\caption{The violin plots of the Dice scores for 132 segmented brain regions using FAST-AID Brain on different inter-domain test sets, i.e., Colin27, HarP, Hammers, and IBSR. The three plots from the top represent the asymmetric, left, and right regions of the brain, respectively. The encompassed white circles show the median points on the violin plots, and the transparent areas visualize the kernel density plots or distributions of the scattered points.}
\label{fig_fastaid_inter}
\end{figure*}

\begin{figure*}[!t]
\centering
\begin{subfigure}[t]{0.98\textwidth}
\raisebox{-\height}{\includegraphics[scale=0.3]{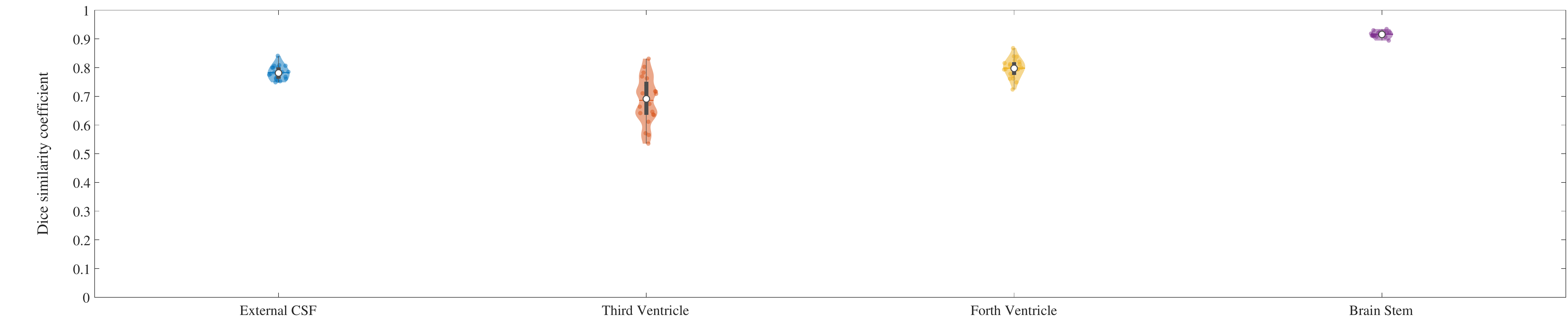}}
\end{subfigure}
\begin{subfigure}[t]{0.98\textwidth}
\raisebox{-\height}{\includegraphics[scale=0.3]{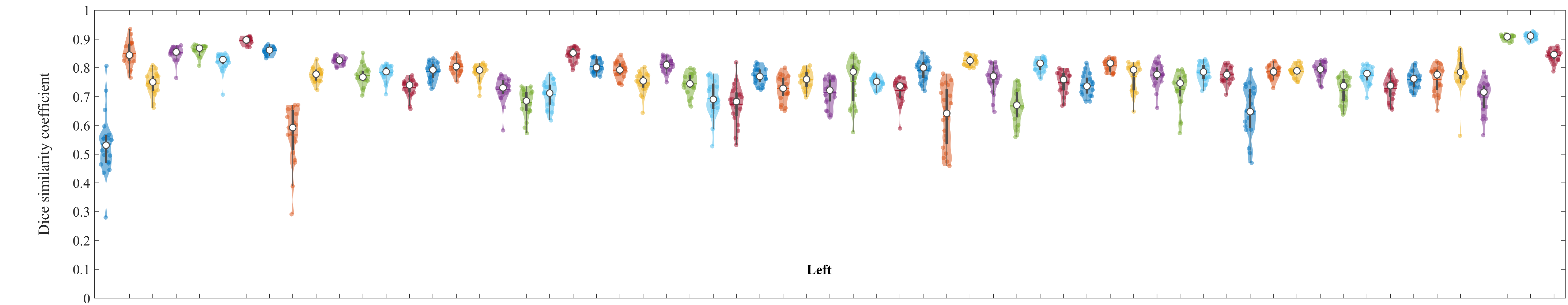}}
\end{subfigure}
\begin{subfigure}[t]{0.98\textwidth}
\raisebox{-\height}{\includegraphics[scale=0.3]{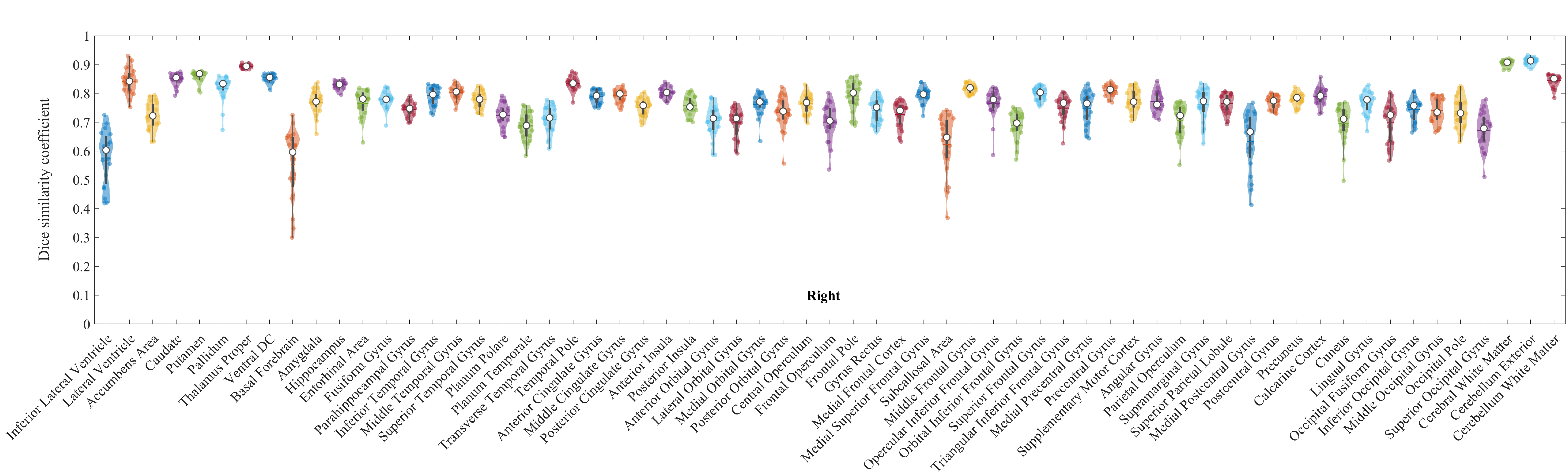}}
\end{subfigure}
\caption{The violin plots of the Dice scores for 130 manually segmented brain regions using the two registered scans of each subject from 20Repeats. The three plots from the top represent the asymmetric, left, and right regions of the brain, respectively. The encompassed white circles show the median points on the violin plots, and the transparent areas visualize the kernel density plots or distributions of the scattered points.}
\label{fig_rescan}
\end{figure*}

To further inspect the regional precision of the proposed segmentation tool, we display the violin plots of DSCs for all segmented brain regions using different intra-domain test sets in Figure \ref{fig_fastaid_intra}. As it can be deduced from the plots, the method performs almost similarly on the left and right compartments, and it shows a very robust high accuracy in several regions including but not limited to the brain stem, ventricles, hippocampus, thalamus, insula, and white matter. Still, the accuracy is relatively low in a few regions of interest such as the basal forebrain.

\subsection{Inter-Domain Generalization}

We also assess the segmentation accuracy of the trained models on the test sets from different domains annotated differently. Table \ref{table3} compares the segmentation accuracy of different methods on the common regions of the annotated test sets. As it can be seen, our proposed model achieves the best results in most of the cases compared to FastSurfer. However, FastSurfer obtains a higher accuracy on HarP, most likely due to training on more samples from the ADNI data with fewer regions for segmentation. In return, FAST-AID Brain attains a more robust accuracy across all inter-domain test sets, higher volumetric similarity, and better results on the higher resolution scan of Colin27. The higher volumetric similarity alongside the higher Dice score on the test sets from different annotation protocols and acquisition parameters can refer to more robust and accurate segmentation results. Nevertheless, the label disagreements and differences in the quality and size of the annotated regions and scans of Hammers compared to the ones used to train the models result in a DSC drop in both models.

To further inspect the regional precision of the proposed segmentation tool, we display the violin plots of DSCs for all segmented brain regions using different inter-domain test sets in Figure \ref{fig_fastaid_inter}. As it can be deduced from the plots, the method does not perform similarly on the left and right compartments, and it shows a somewhat robust high accuracy in some regions such as the ventricles and white matter. It should be noted that besides the domain shift problem the DSC drop and instabilities could be because there is no exact matching between the labels from different annotated datasets and those of the trained models.

\subsection{Segmentation Goodness}

In previous sections, we used two similarity metrics (DSC and VS) to evaluate the segmentation accuracy of the models by comparing the results of automatic segmentation with the available ground truth labels. Apart from the automatic segmentation accuracy, the goodness of the obtained values could be affected by the ground truth annotations due to differences in the anatomies (e.g., size, shape, border), artifacts, labeling protocols, and the precision of manual segmentation. Some regions of interest such as ventricles are relatively easy to segment, as their borders are well-defined, whereas some other regions such as the amygdala or cingulate are difficult, as their boundaries are ambiguous. To see these effects, we use the 20Repeats dataset where 20 subjects from the OASIS were scanned twice by some time laps and the two scans were manually labeled \citep{Worth2015} and registered to calculate DSC and VS for the corresponding cortical and subcortical regions. The manual segmentation could achieve DSC of $0.842\pm0.021$ and VS of $0.946\pm0.033$ on the rescanned images from the 20Repeats dataset. Although the obtained DSCs of the matched manual segmentations are higher than the automatic segmentation results with a DSC of $0.788\pm0.085$, the calculated volumes are comparable with those of the automatic segmentation with a VS of $0.941\pm0.062$. More interestingly, the estimated values indicate that there is on average around 11\% difference in the manually segmented volumes.

To be more specific, Figure \ref{fig_rescan} shows the violin plots of DSCs for all manually segmented brain regions using the two scans of each subject from 20Repeats. As can be seen, there are always disagreements in neuroanatomical labeling systems. They cannot achieve a DSC of 1 in segmenting almost the same images and can see a drop in DSC (e.g., less than 0.3) in some regions (e.g., basal forebrain). Moreover, the manual segmentation accuracy is in line with the results of the automatic segmentation shown in Figure \ref{fig_fastaid_intra} and Figure \ref{fig_fastaid_inter}, indicating that the automatic model follows the human patterns in segmenting different areas of the brain. In other words, one can see the same behavior in the corresponding regions of manually labeled scans and those of the automatic segmentation, e.g., the DSC drop in the basal forebrain.

\subsection{ICV Estimation and Brain Development}

\begin{figure*}[!t]
\centering
\begin{subfigure}[t]{0.475\textwidth}
\raisebox{-\height}{\includegraphics[scale=0.6]{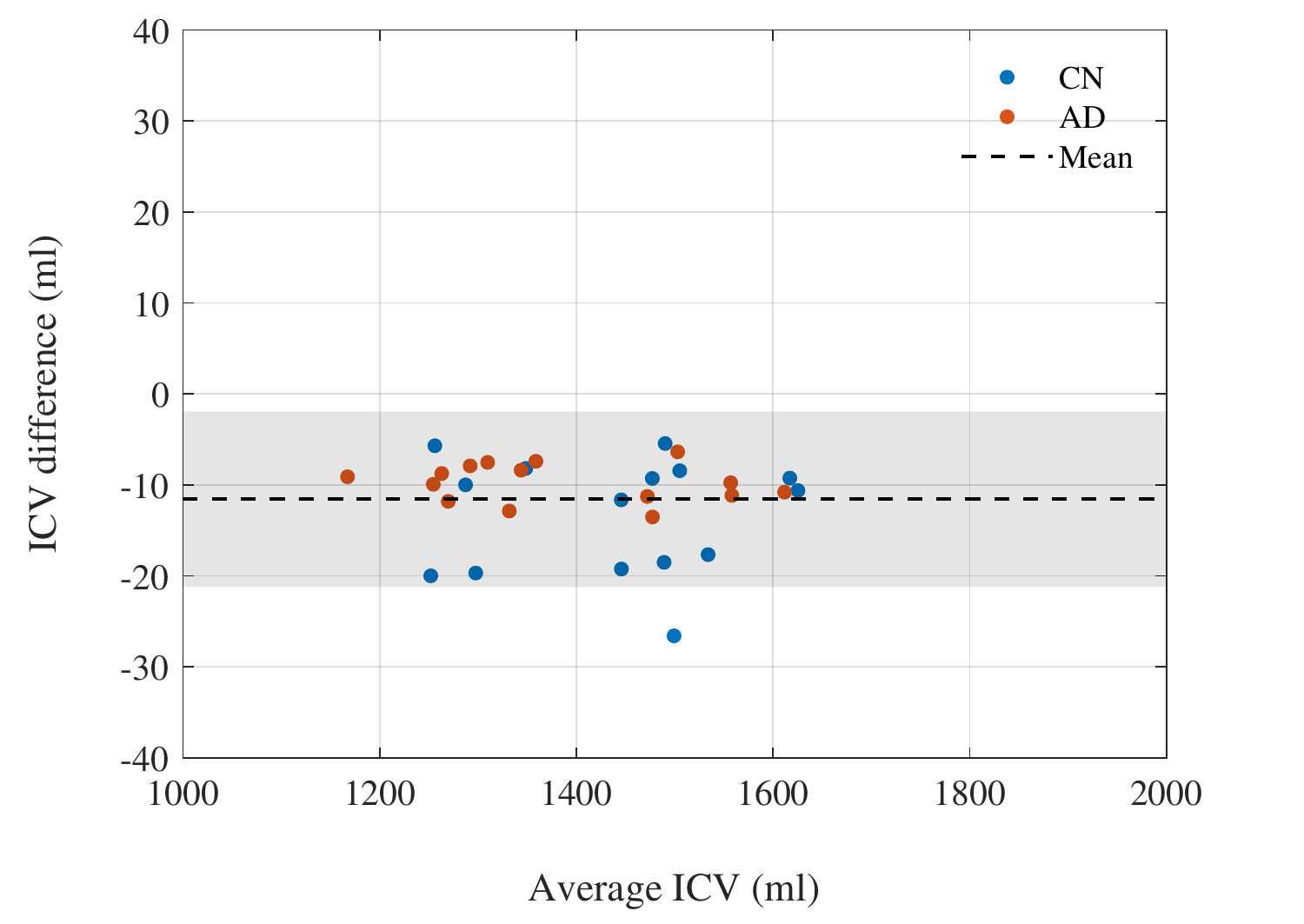}}
\end{subfigure}
\begin{subfigure}[t]{0.475\textwidth}
\raisebox{-\height}{\includegraphics[scale=0.6]{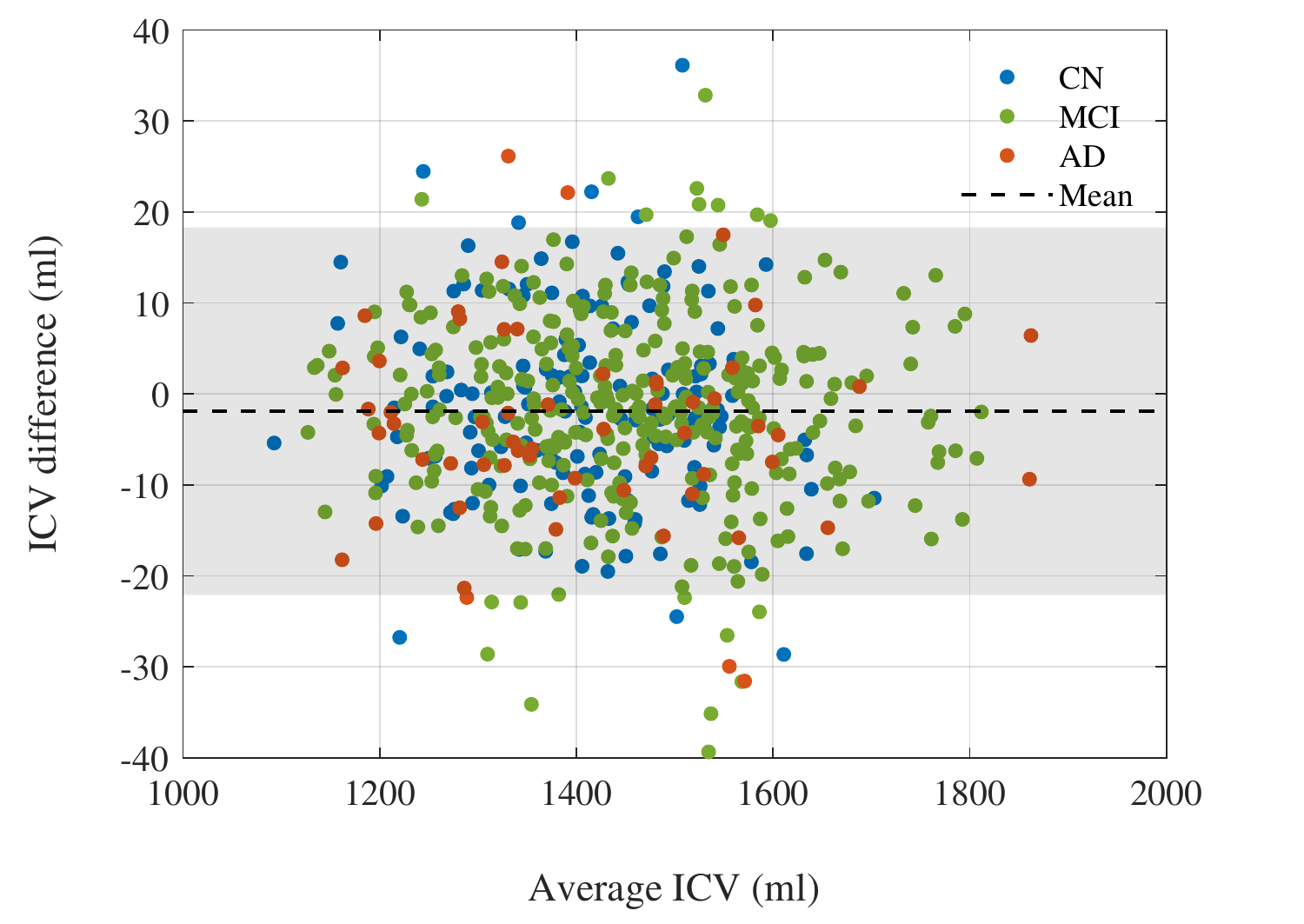}}
\end{subfigure}
\caption{The Bland-Altman plots for ICV differences versus the average with 95\% limits of agreement around the mean. Left: Estimates obtained based on the segmented scans from the ADNI30 using the predicted and true labels. Right: Estimates obtained based on the segmented scans from the ADNI1-2Yr using the predicted follow-up and predicted baseline labels.}
\label{fig_fastaid_icv}
\end{figure*}

The proposed segmentation tool can be used to estimate the ICV and capture the volumetric changes over time. First, we predict the whole intracranial labels (cranial cavity and all brain compartments) for the scans from ADNI30 and use the aggregated label volumes as the estimated ICV for the automatically segmented brain scans. The left subfigure of Figure \ref{fig_fastaid_icv} shows the Bland-Altman plot \citep{Bland1999} for ICV differences (automatic - manual) versus the average (automatic and manual) with 95\% limits of agreement around the mean. The obtained estimated ICV differences using the proposed segmentation model are very small compared to the state-of-the-art results \citep{Malone2015,Sargolzaei2015} which include SPM12, FSL, and FreeSurfer.

As a complementary experiment, we use the obtained model to segment MRI scans from yearly follow-ups of the ADNI1-2Yr subjects. The main purpose is to see how cognitively normal (CN), mild cognitive impairment (MCI), and Alzheimer’s disease (AD) subjects develop in the course of AD using the acquired regional segments and to evaluate the stability of the model on the data with no manual annotations based on the estimated volumes within different groups. The right subfigure of Fig \ref{fig_fastaid_icv} shows the Bland-Altman plot for ICV differences (follow-up - baseline) versus the average (follow-up and baseline) with 95\% limits of agreement around the mean. As can be seen, the estimated yearly ICV differences for different groups stay small, while there are significant changes over time in the regional volumes of the elderly subjects.

\begin{figure*}[!t]
\centering
\begin{subfigure}[t]{0.98\textwidth}
\raisebox{-\height}{\includegraphics[scale=0.31]{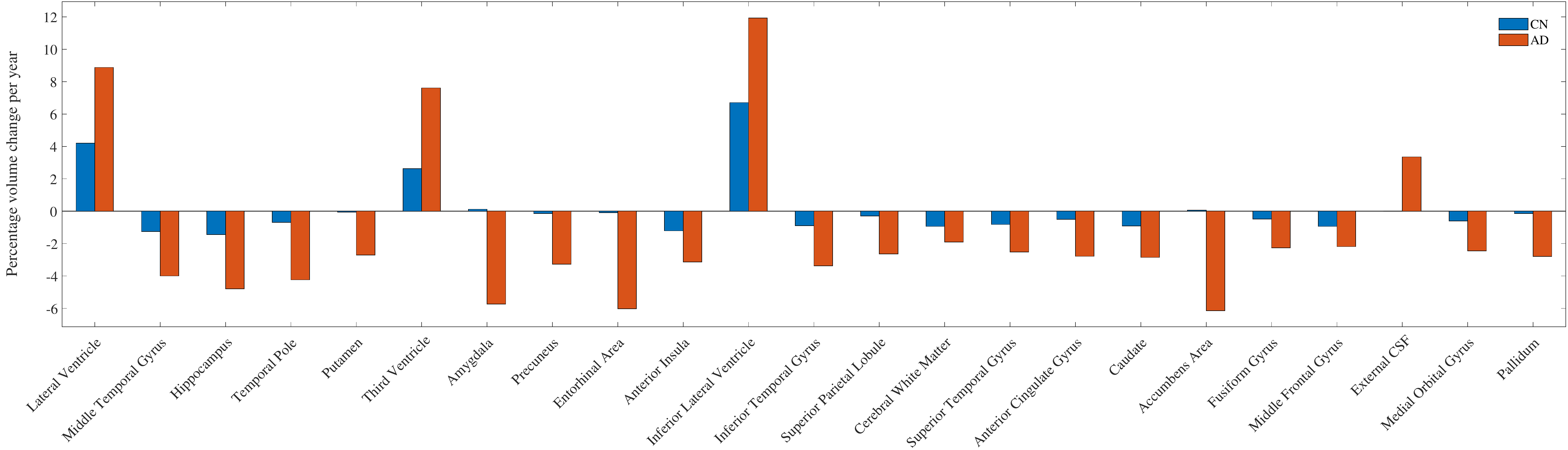}}
\end{subfigure}
\caption{The annual percentage changes in regional volumes obtained using the segmented scans from the ADNI1-2Yr sorted by the significance of the difference between the normal and demented groups.}
\label{fig_changes}
\end{figure*}

To see how the segmented regions of the two groups are statistically significantly different, we apply the two-sided Wilcoxon rank-sum or Mann-Whitney U test \citep{Mann1947} to the obtained yearly volume changes of the patients. Figure \ref{fig_changes} shows the annual percentage volume changes of the regions with significant difference ($p < 0.05$) between the groups. The obtained results are in line with the literature findings \citep{Pai2013}, where the lateral ventricle and hippocampus have the most significant changes compared to the other compartments in the course of AD.

\section{Conclusion}

In this study, we proposed a novel deep learning method for automatic segmentation of the human brain into 132 regions using an efficient 2.5D U-Net-like network applied to the three principal views. The proposed model benefitted from the intersection points of different views and hierarchical relations for the fusion during the end-to-end training. Weak supervision was used to learn from partially labeled data to segment the whole brain and estimate the ICV, and data augmentation was employed in the training step to expand the data with realistic artifacts and variations for robust training of the model while preserving data privacy.

Several experiments using different atlases were conducted to evaluate the segmentation performance of the trained model compared to the state-of-the-art. The results indicated that the proposed model was accurate and robust to domain shifts compared to the existing methods \citep{Henschel2020}, in terms of both volumetric and Dice similarity, when applied to different intra- and inter-domain datasets. The average inference time for segmenting an MRI scan was less than 40 seconds using an NVIDIA GeForce RTX 2070 GPU machine with 8 GB memory. The proposed tool is also made publicly available at \url{https://github.com/Mostafa-Ghazi/FAST-AID-Brain}.

Although the intra-domain results are comparable with the state-of-the-art \citep{Huo2019}, the proposed tool is extremely fast, uses a limited number of scans for training, and can achieve higher generalization accuracy on data from different domains, thanks to efficient data augmentation techniques used in training and avoiding preprocessing and postprocessing techniques such as registration, skull stripping, and bias field correction. The applied methods helped to generate realistic brain data with high variability and make the model robust to domain shift, especially in the ADNI case where the scans could have lower contrast and contain larger areas and more artifacts beyond the head. The proposed model is developed as a tool with very few dependencies, which makes it suitable for real applications.

The proposed approach for weak supervision of the voxels with missing labels, i.e., cranial cavities, helped with an accurate estimation of the ICV compared to the state-of-the-art \citep{Malone2015,Sargolzaei2015,Liu2022}. This measure is typically used for the normalization of the regional brain volumes for head size in studies associated with brain volume changes such as AD, where inaccurate estimation of ICV can introduce bias in the outcome. Note that the ICV defined in this study includes the whole brain volume, i.e., brainstem, infundibular and pituitary, cerebellar, subcortical, and cerebral parenchyma, and total intracranial CSF in ventricular and subarachnoid spaces, excluding skull, subcutaneous and orbital fat, mastoid and nasal sinuses, dural venous sinuses, larger blood vessels beyond the brain surface, bony protuberances (dorsum sellae), and cranial nerve roots.

Last but not least, the accuracy of the automatic segmentation could be affected by manual segmentation and the complexity of the regions. A simple scan-rescan segmentation experiment showed that a DSC of 1 may not be achievable by humans, there could be regions with DSCs below 0.3, and there could be on average around 11\% difference in the manually segmented volumes. Still, automatic segmentation in practice follows the same accuracy patterns and uncertainties in segmenting different regions.

\section*{Disclosures}

M. Nielsen is shareholder in Biomediq A/S and Cerebriu A/S.

\section*{Acknowledgments}

This project has received funding from the European Union's Horizon 2020 research and innovation programme under the Marie Sk{\l}odowska-Curie grant agreement No. 643417, No. 681043 and No. 825664, VELUX FONDEN and Innovation Fund Denmark under the grant number 9084-00018B, and Pioneer Centre for AI, Danish National Research Foundation, grant number P1.

Data collection and sharing for this project was funded by the Alzheimer's Disease Neuroimaging Initiative (ADNI) (National Institutes of Health Grant U01 AG024904) and DOD ADNI (Department of Defense award number W81XWH-12-2-0012). ADNI is funded by the National Institute on Aging, the National Institute of Biomedical Imaging and Bioengineering, and through generous contributions from the following: AbbVie, Alzheimer's Association; Alzheimer's Drug Discovery Foundation; Araclon Biotech; BioClinica, Inc.; Biogen; Bristol-Myers Squibb Company; CereSpir, Inc.; Cogstate; Eisai Inc.; Elan Pharmaceuticals, Inc.; Eli Lilly and Company; EuroImmun; F. Hoffmann-La Roche Ltd. and its affiliated company Genentech, Inc.; Fujirebio; GE Healthcare; IXICO Ltd.; Janssen Alzheimer Immunotherapy Research \& Development, LLC.; Johnson \& Johnson Pharmaceutical Research \& Development LLC.; Lumosity; Lundbeck; Merck \& Co., Inc.; Meso Scale Diagnostics, LLC.; NeuroRx Research; Neurotrack Technologies; Novartis Pharmaceuticals Corporation; Pfizer Inc.; Piramal Imaging; Servier; Takeda Pharmaceutical Company; and Transition Therapeutics. The Canadian Institutes of Health Research is providing funds to support ADNI clinical sites in Canada. Private sector contributions are facilitated by the Foundation for the National Institutes of Health (www.fnih.org). The grantee organization is the Northern California Institute for Research and Education, and the study is coordinated by the Alzheimer's Therapeutic Research Institute at the University of Southern California. ADNI data are disseminated by the Laboratory for Neuro Imaging at the University of Southern California.

\bibliographystyle{model2-names.bst}\biboptions{authoryear}
\bibliography{references}

\end{document}